\documentclass[aps,prx,twocolumn,nopreprintnumbers,noeprint,superscriptaddress]{revtex4-2}
\usepackage{graphicx}
\usepackage{dcolumn}
\usepackage{bm}
\usepackage{amsmath}
\usepackage{amsfonts}
\usepackage{psfrag}
\usepackage{graphicx}
\usepackage{color} 
\usepackage{footmisc}
\usepackage{mathtools}
\usepackage{amssymb}
\usepackage[normalem]{ulem}
\usepackage[english]{babel}
\usepackage[utf8]{inputenc}
\usepackage{hyperref}
\hypersetup{%
	colorlinks = true,
	citecolor = blue,
	linkcolor = blue,
	urlcolor = blue,
	linkbordercolor = 0 0 0,
	pdfborder= 0 0 0,
}

\begin{abstract}
Recent experimental advances in Positronium (Ps) physics have made it possible to produce dense Ps ensembles in which Ps-Ps interactions may occur, leading to the production of Ps$_2$ molecules and paving the way to the realization of a Ps Bose-Einstein Condensate (BEC). In order to achieve this latter goal it would be advantageous to develop new methods to measure Ps densities in real-time. Here we describe a possible approach to do this using polaritonic methods: using realistic experimental parameters we demonstrate that a dense Ps gas can be strongly coupled to the photonic field of a distributed Bragg reflector microcavity. In this strongly coupled regime, the optical spectrum of the system is composed of two hybrid positronium-polariton resonances separated by the vacuum Rabi splitting, which is proportional to the square root of the Ps density.
Given that polaritons can be created on a sub-cycle timescale, a spectroscopic measurement of the vacuum Rabi splitting could be used as an ultra-fast Ps density measurement in regimes relevant to Ps BEC formation. Moreover, we show how positronium-polaritons could potentially enter the ultrastrong light-matter coupling regime, introducing a novel platform to explore its non-perturbative phenomenology.
\end{abstract}

\begin{document}
 
\title{Positronium density measurements using polaritonic effects}

\author{Erika Cortese}
\affiliation{School of Physics and Astronomy, University of Southampton, Southampton, SO17 1BJ, United Kingdom}

\author{David B. Cassidy}
\affiliation{Department of Physics and Astronomy, University College London, Gower Street, London WC1E 6BT, United Kingdom}

\author{Simone De Liberato}
\affiliation{School of Physics and Astronomy, University of Southampton, Southampton, SO17 1BJ, United Kingdom}
\email{E-mail: s.de-liberato@soton.ac.uk}

\maketitle

\section{Introduction}
\label{sec:intro}
 
Positronium (Ps), the electron-positron bound state, is a meta-stable two-body atomic system that has a lifetime against self-annihilation of 142~(0.125)~ns in the triplet (singlet) ground state~\cite{Bethe1957}. Since Ps is composed only of leptons it is almost fully described by bound-state quantum electrodynamics (QED)~\cite{karshenboim_precision_2004}, and can therefore be used to test QED theory via precision measurements of Ps energy levels or decay rates~\cite{adkins_precision_2022}. 
 
The existence of Ps atoms was first suggested by {Mohorovi{\v c}i{\' c} in 1934~\cite{Mohorovicic1934}, with subsequent, independent, predictions by Pirenne~\cite{pirenne_proper_1946}, Ruark~\cite{Ruark1945} and Wheeler~\cite{Wheeler1946}. Wheeler also considered what he called polyelectrons, which are systems containing more than one electron and/or positron, the simplest case being the Ps atom. He showed that three-body Ps ions, comprising two electrons and one positron (or two positrons and one electron), would also form meta-stable bound states. Although Wheeler was unable to determine if four-body Ps$_2$ molecules would be stable, this was subsequently shown to be the case by Hylleraas and Ore~\cite{Hylleraas1946}.  
 
Ps atoms were first produced experimentally in 1951 by Deutsch using a gas cell apparatus~\cite{deutsch_evidence_1951}.  The development of slow positron beams~\cite{Coleman2000} in the 1970's allowed for more controlled Ps production using solid surfaces~\cite{Schultz1988}, and later also the production of the negative Ps ion~\cite{Mills1981} and Ps$_2$ molecules~\cite{cassidy_production_2007}. 
In addition to creating polyelectrons, a long term goal of Ps physics has been the formation of an ensemble of Ps atoms that are cold/dense enough to create a Bose-Einstein Condensate (BEC)~\cite{Platzman1994}. The primary motivation for producing a Ps BEC is that such a system may exhibit the phenomenon of stimulated annihilation~\cite{Bertolotti1979, Liang1988}, allowing for the creation of a gamma-ray laser, but the properties of a Ps BEC are also of interest from a theoretical perspective (e.g.,~\cite{Froelich2006, Wang2014, Avetissian2014, Yunjin2021}). The low mass of Ps means that the transition temperature (that is, the temperature at which a dense ensemble will undergo a phase transition to form a condensate) is considerably higher than it is for all other atoms; for example room temperature Ps condensates could form at densities on the order of 10$^{20}$~cm$^{-3}$~\cite{cassidy_experimental_2018}. Since Ps can be cooled via collisions to ambient cryogenic temperatures in microcavities~\cite{Shu2021} this density requirement may be reduced to the 10$^{19}$~cm$^{-3}$ level for experimentally accessible temperatures. The specifics of various Ps production methods may also allow for significant density enhancements (e.g., \cite{mills_positronium_2019, asaro_conditions_2022}). 

It is evident that any practical scheme designed to produce a Ps BEC requires a high density positron beam, and an efficient means to generate a correspondingly high Ps density. Recent advances in positron trapping and control methods~\cite{Danielson2015} have made it possible to produce Ps in porous silica films at densities that allow for Ps-Ps interactions to occur~\cite{Cassidy2008b}, resulting in the formation of a spin polarized Ps gas with an average density on the order of 10$^{16}$~cm$^{-3}$~\cite{Cassidy2011}, with higher Ps densities expected in the future~\cite{Mills2019}. 

The optimization of any experimental schemes to generate a Ps BEC would benefit from direct measurements of the Ps temperature and density. However, these are not trivial measurements: Ps temperatures can be directly measured using the angular correlation of annihilation radiation~\cite{Cecchini2018}, but this requires the atoms to be magnetically quenched and decay via a two-photon annihilation process. The rate of annihilation events following Ps-Ps scattering can be measured using single-shot lifetime methods~\cite{Cassidy2006}, from which one may infer the Ps density. However, once the Ps gas has become fully spin-polarized this method no longer provides any signal~\cite{cassidy_production_2010}. Another way to determine Ps densities would be to measure the density-dependant collisional frequency shift~\cite{higgins_low-energy_2019} of Ps atomic transitions. This technique has not been demonstrated for confined Ps, and may be affected by the detailed interactions between the Ps gas and its environment~\cite{gurung_resonant_2020}. 

In this paper we describe an alternative approach to this problem that exploits concepts and techniques of cavity quantum electrodynamics (CQED), the field that investigates the interaction between dipolar active transitions in atoms, molecules or other materials, and single photons inside an optical cavity~\cite{Walther2006}. When the light-matter coupling strength, referred to as Vacuum Rabi Frequency (VRF), becomes larger than the loss rates of the light and matter excitations, the system enters the so-called \emph{strong coupling} (SC) regime \cite{thompson_observation_1992, weisbuch_observation_1992}. In this regime the physics of the system can be correctly described only in terms of the light–matter hybrid eigenmodes of the coupled system, often named \emph{polaritons} \cite{ballarini_polaritonics_2019, basov_polariton_2021}, which manifest themselves as spectral resonances split by twice the VRF. The collective nature of light-matter coupling inside the cavity leads to a VRF splitting that scales as the square root of the number of optically active dipoles within the cavity mode volume \cite{dicke_coherence_1954}. The possibility of achieving SC between light and collective matter excitations, and thus to observe polaritons in a variety of solid-state systems, has led to the achievement of landmarks as room-temperature BEC~\cite{kasprzak_boseeinstein_2006} and room-temperature superfluidity~\cite{lerario_room-temperature_2017}. 

\begin{figure}[tp]
    \centering
    \includegraphics[width=0.48\textwidth]{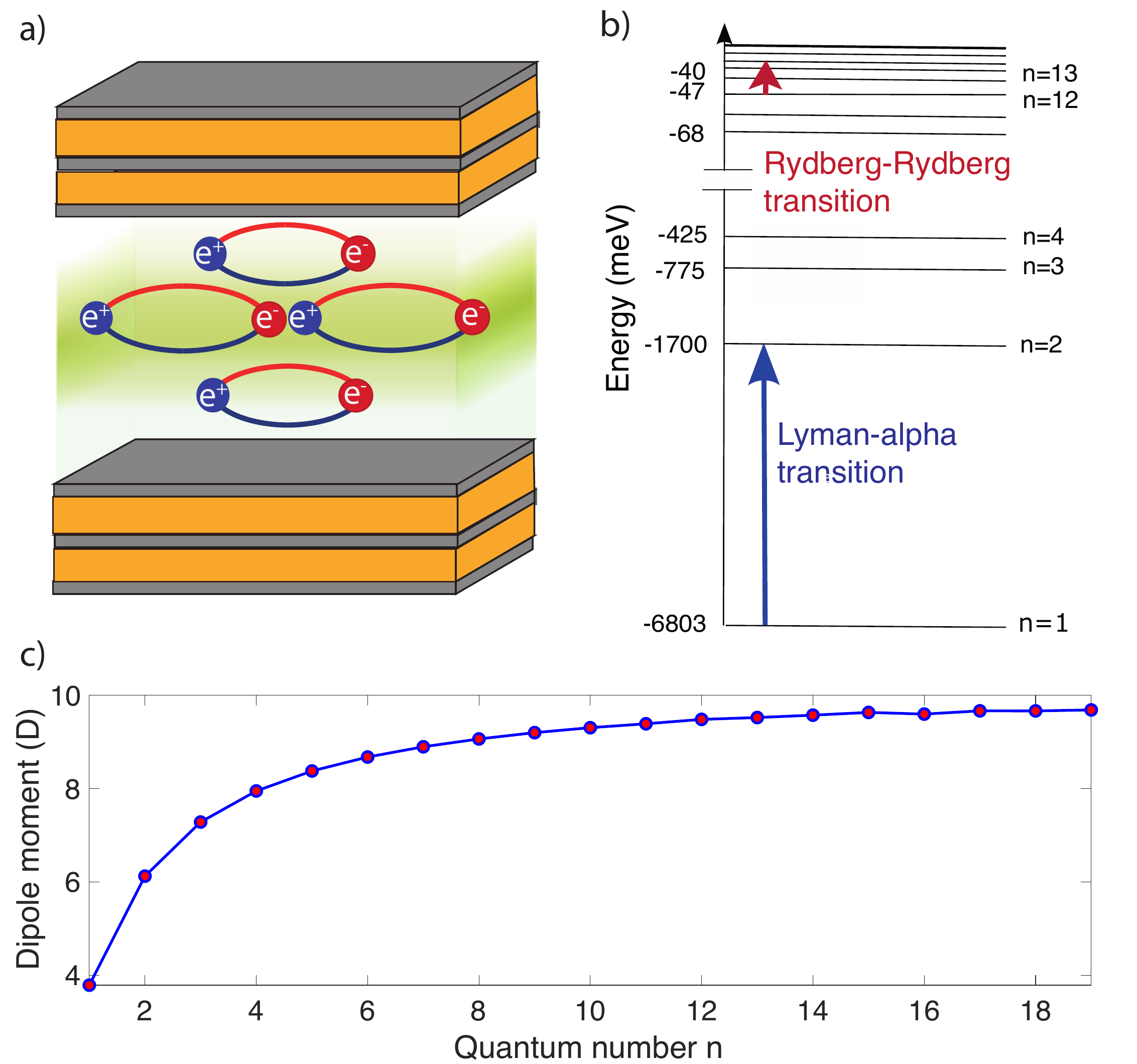}
    \caption{(a) Sketch of Ps CQED system, representing Ps atoms interacting with the electromagnetic field trapped in a cavity formed by two DBRs. (b) Energy levels of Ps with highlighted the Lyman-alpha and Rydberg-Rydberg transitions explicitly considered in this paper (not to scale). (c) Dipole moment $d_{n\rightarrow n+1}$ in Debyes as a function of the principal quantum number $n$ of the initial energy state.}
    \label{fig1}
\end{figure}

Here we theoretically investigate the light-matter interaction between the electronic transitions of a collection of Ps atoms and the confined electromagnetic field of an optical cavity, and explore the conditions under which such a system can access the SC regime where it would be possible to observe positronium-polariton modes. Since the cavity-induced splitting in the SC regime is proportional to the square root of the density of the atoms, producing a Ps ensemble inside an optical cavity
would lead us to a direct measurement of the Ps density via spectroscopic measurements. Moreover, since the formation of polaritons can occur on sub-period time scale \cite{gunter_sub-cycle_2009}, this mechanism can provide an ultrafast measurement of the Ps density, allowing for real-time measurements even under conditions where the Ps density is varying.   
 
 Distributed Bragg Reflectors (DBRs), cavities composed of a pair of mirrors composed of alternating layers of semiconductor materials, are a flexible and successful resonator technology widely used in polaritonics~\cite{kavokin_microcavities_2017,ballarini_polaritonics_2019}. Although usually covering the near-infrared and visible ranges, DBRs can also be effective at shorter and longer wavelengths, and thus can be compatible with a wide variety of Ps transitions. Deep-UV DBRs have been recently fabricated via deposition of AlGaN/AlN alternating layers with very promising Q-factors~\cite{detchprohm_sub_2017}. For mid- to far-infrared ranges, DBRs consisting of silicon slabs in air have been used~\cite{li_vacuum_2018}, with high quality factors obtained owing to a large refractive index mismatch. 
 
 Open cavity systems have been largely employed to overcome the issue of growing the active material on top of the mirrors, allowing both non-invasive investigation and good control over the microcavity resonances and mode volumes~\cite{dufferwiel_excitonpolaritons_2015, betzold_tunable_2018}. These systems are particularly suitable to the present investigation as different kinds of microscopic Ps producing materials, such as porous silica layers~\cite{Cassidy2010} or MgO powders~\cite{gurung_positronium_2019}, can be placed upon a prefabricated DBR with different thickness and surface extension [see Fig.~\ref{fig1}~(a)].

\section{Light-matter coupling in positronium}

\subsection{Theoretical framework}
\label{sec:theory}
At the Bohr level, the energy levels of Ps atoms are formally equivalent to those of the hydrogen atom, with reduced mass $\mu_{\mathrm{Ps}} = m_e/2$ and Bohr radius $a_{\mathrm{Ps}} = 2 a_0$, where $m_e$ is the electron mass and $a_0$ the Bohr radius. Fine structure corrections are considerably different in Ps~\cite{adkins_precision_2022} but these difference are unimportant in the present case. The Bohr energy spectrum, sketched in Fig.~\ref{fig1}~(b), similar to that for the hydrogen atom, becomes
\begin{eqnarray}
    E_n=-\frac{\text{Ry}}{2n^2} \,\textrm{eV},
\end{eqnarray}
where $n$ refers to the principal quantum number, and the Rydberg constant $\text{Ry}\approx 13.605$~eV. We will consider transitions between states of the form $| n S \rangle$ and  $| (n+1) P \rangle$, corresponding to energies $\hbar\omega_{n\rightarrow n+1}= E_{n+1}-E_n$ and dipoles
$d_{n\rightarrow n+1}=-e\langle (n+1) P | z | n S \rangle$. We choose the phases to make the dipoles real and positive [see Fig. \ref{fig1}(c)].

The interaction between the optically active transition of a collection of $N$ Ps atoms and the single-mode cavity field oscillating at frequency $\omega_c$, can be described in the dipolar gauge \cite{de_bernardis_breakdown_2018} by a VRF of the form
\begin{eqnarray} \label{OR}
   \Omega_{n\rightarrow n+1}&=&\sqrt{\frac{ N\hbar \omega_c}{2 \epsilon_0 \epsilon   S L_{\textrm{eff}}}} d_{n\rightarrow n+1},
\end{eqnarray}
 where $\epsilon$ is the dielectric constant of the medium in which the Ps atoms are trapped, $V=S  L_{\textrm{eff}}$ is the mode volume, estimated as the product of the effective cavity mode length $L_{\textrm{eff}}$ and the area $S$ occupied by the atoms. For $\lambda/2$ DBR cavities the effective length is $L_{\textrm{eff}}=L_c+L_{\textrm{DBR}} $ where $L_c=\lambda_c/2 \sqrt{\epsilon}$, with mode wavelength $\lambda_{c}$, and $L_{\textrm{DBR}}$ is the DBR penetration depth \cite{kavokin_microcavities_2017}.
 
In the following we report the results of the numerical simulations for two distinct transitions, involving very different setups and energy scales. The first (A) ($n=1 \rightarrow 2$) is the Lyman-alpha transition at energy $\hbar\omega_{A}=5.111$ eV, which includes the slight blue shift due to the pore trapping observed in standard porous silica systems \cite{cassidy_cavity_2011}. The second (B) ($n=12 \rightarrow 13$) is a Rydberg-Rydberg transition at energy $\hbar \omega_{B}=7$ meV.

\subsection{Lyman-alpha transition ($n=1 \rightarrow 2$)}
\label{subsec:1S}
In order to cover the $\hbar\omega_{A}=5.111$ eV Lyman alpha transition we consider a resonant AlGaN/AlN DBRs cavity, sketched in Fig.~\ref{fig2}~(a)~\cite{detchprohm_sub_2017}, with mirrors made of $r=30$ repetitions of a pair of layers with thicknesses $l_1=22.7$ nm (AlGaN) and $l_2=25.2$ nm (AlN). Given the refractive indexes $n_1=2.67$ (AlGaN) and $n_2=2.42$ (AlN) this choice allows us to respect the Bragg condition 
\begin{eqnarray} \label{Bragg}
n_1 l_1=n_2 l_2= \lambda_{c}/4,
\end{eqnarray}
with $\lambda_{c}=2\pi c/\omega_{A}$ considered at the resonance point.

Lifetimes for the singlet $1^1S_0$ and the triplet $1^3 S_1$ states have been calculated, and measured by QED experiments \cite{adkins_precision_2022} and are $125$~ps and $142$~ns respectively. Although the decay rate of a free Ps atom in a particular spin-state is well-defined, it may be substantially different for atoms interacting with external surfaces. Factors determining extrinsic properties of Ps atoms will be here all enclosed in the phenomenological lifetimes observed on the experimental Ps Lyman transition spectra \cite{cassidy_cavity_2011}, corresponding to a FWHM linewidth $\hbar\gamma_A=0.64$ meV, likely dominated by Doppler broadening. Note that the large broadening reported in Ref.~\cite{cassidy_cavity_2011} might be reduced in other samples due Dicke-narrowing \cite{dicke_effect_1953}.

\begin{figure*}[tb]
    \centering
    \includegraphics[width=\textwidth]{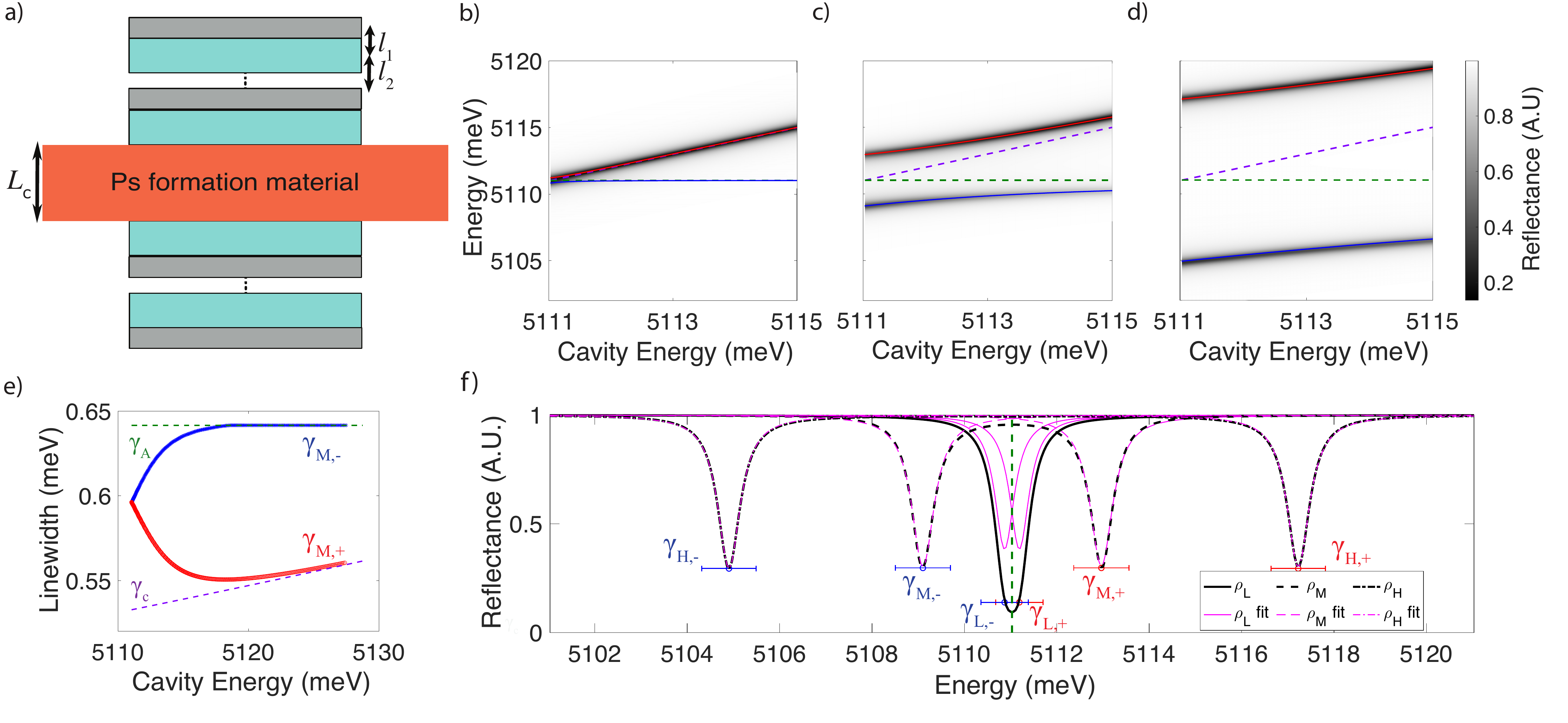}
    \caption{(a) Sketch of a proposed experimental set-up, composed by a AlGaN/AlN DBRs composed of $r=30$ repetitions embedding a Ps formation material. (b-d) Reflectance spectra of the cavity-Ps system as function of the bare cavity frequency (violet dashed line), calculated via transfer matrix at the Ps atoms density $\rho_L=10^{16}$ cm$^{-3}$ (b), $\rho_M= 10^{18}$ cm$^{-3}$ (c) and $\rho_H=10^{19}$ cm$^{-3}$ (d). The bare transition energy is marked in all the panels by a green dashed line and the calculated polariton modes are blue (lower polariton) and red (upper polariton) solid curves. (e) Plots of polariton linewidths $\hbar\gamma_{M,\pm}$ at density $\rho_M$ as function of the cavity energy (blue stars for lower polariton and red dots for upper polariton) compared to the bare cavity (violet dashed line) and bare matter loss (green dashed line).
    (f) Reflectance spectra at the resonance condition $\omega_c=\omega_A$ (black dashed line). The polaritonic resonances have been fitted by Lorentzian functions (magenta lines) from which the linewidths of the different resonances have been extracted.}
    \label{fig2}
\end{figure*}

Panels (b-d) of Fig. \ref{fig2} show the reflectance spectra calculated by transfer matrix method (see Sec. \ref{sec:TM}) as a function of the bare cavity mode $\hbar \omega_c$ (violet dashed line), tuned by variation of the incident angle in proximity of the resonance point $\hbar \omega_{A}$ (green dashed line) for three different Ps densities: (b)~$\rho_L=10^{16}$ cm$^{-3}$, corresponding to the present experimental record \cite{cassidy_production_2010}, (c)~$\rho_M= 10^{18}$ cm$^{-3}$ and (d)~$\rho_H=10^{19}$ cm$^{-3}$, corresponding to the approximate value expected to be required for BEC formation~\cite{asaro_conditions_2022}. 

The coupled light-matter polariton modes, the upper one marked with blue and the lower one with red solid curves have been calculated by diagonalising a bosonic Hopfield Hamiltonian in the rotating wave approximation (RWA) (see Sec. \ref{sec:bosonic}). These curves have been plotted on top of the reflectance data from the transfer matrix calculations in panels (b-d), perfectly matching them. 
Fig. \ref{fig2}(f) displays the reflectance spectra (black curves) at the resonance $\omega_{c}=\omega_{A}$ for the same three densities, with resonances fitted by a pair of Lorentzian functions (magenta curves), from which we can extract the linewidth $\gamma_{j,\pm}$ of the lower (-) and upper (+) polariton modes for each density $j=\lbrack L,M,H\rbrack$.
In panel (e) we plot the linewidths for the two polariton branches for the medium density, as well as the bare linewidths of the matter ($\gamma_A$) and photonic ($\gamma_c$) resonances, showing how the polaritonic linewidth is matter-limited, i.e., improving the DBRs quality factor will lead to only a marginal improvement toward the achievement of SC.

Overall we see that in the case of low density the condition of SC is not achieved and we are unable to resolve the resonant polaritonic splitting, but that such a splitting becomes visible at densities which, albeit larger, remain more than one order of magnitude smaller of what expected to be necessary to achieve BEC. This technique is, therefore, well-suited as a diagnostic tool for the optimization of such experiments.

\subsection{Rydberg-Rydberg transitions ($n=12 \rightarrow 13$)}

In order to address the $\hbar\omega_{B}=7$ meV transition we consider the same high-Q photonic crystal cavity designed in Ref~\cite{li_vacuum_2018}. The resonator, sketched in Fig.~\ref{fig4}~(a), consists of two pairs of $13\, \mu$m thick silicon ($n_1=3.4$) slabs in air ($n_2=1$), with spacing between each pair of $44.4\, \mu$m, satisfying Eq.~\ref{Bragg} for $\omega_c\approx \omega_{B}$.

It has been demonstrated that the production of Rydberg Ps in vacuum can occur with an overall efficiency of $\approx$~25\%, determined primarily by the spectral overlap of the UV excitation laser and the Doppler broadened width of the $1S-2P$ transition~\cite{cassidy_efficient_2012}. With the aim of achieving SC, this reduction in the achievable density is more than compensated by the fact that, as shown in Fig.~\ref{fig1}~(c), Rydberg states have larger dipole moments and the corresponding transitions have much smaller linewidths. The intrinsic annihilation rates of Ps atoms depend upon the overlap probability of electron and positron, which scales with $n^{-3}$, and Rydberg states between $n=10$ and $n=15$ maintain a narrow resonance lineshape if compared to the Zeeman-mixed higher-energy states ($n>15$)~\cite{cassidy_efficient_2012}. For the present simulation we use the linewidth observed in the experiments reported in Ref.~\cite{cassidy_efficient_2012}, obtaining $\hbar\gamma_{B}=7.86 \cdot 10^{-5}$ meV. Fig.~\ref{fig4}(b-d) show the polariton resonances at the three values of the Ps density $\tilde{\rho}_{L}=10^{15}$ cm$^{-3}$, $\tilde{\rho}_{M}=10^{17}$~cm$^{-3}$, and $\tilde{\rho}_{H}=10^{18}$ cm$^{-3}$. These have been chosen a factor $10$ smaller than for the Lyman-alpha transition described in SEC~\ref{subsec:1S}, to take into account the efficiency of Rydberg state creation. 

As in the previous case, the calculated polariton modes are shown as blue and red solid curves on top of the spectral resonances. Note that while panel (b) shows a very good agreement between the simulations and the analytical theory, this is less true for panel (c) and not the case at all for panel (d). This is due to the fact that in such a case the resonant VRF becomes a substantial fraction of the bare resonant frequency ($\Omega_{B}/\omega_{B}=0.15$). The system is thus in the ultrastrong coupling regime~\cite{anappara_signatures_2009,frisk_kockum_ultrastrong_2019,forn-diaz_ultrastrong_2019} in which antiresonant terms of the Hamiltonian start to have a non-negligible effect. Although this is beyond the scope of this paper, we notice that such a fact marks positronium-polaritons as an interesting platform in which it could be possible to explore and exploit the rich phenomenology of non-perturbative light-matter coupling, including the presence of virtual excitations in the ground state \cite{de2017virtual} and the possibility of modifying the positronium wavefunction \cite{khurgin_excitonic_2001,brodbeck_experimental_2017}.
Panel (d) also shows further spectral resonances around the bare transition energy.  This is due to a VRF that is large enough to couple the Ps matter excitation to higher DBRs cavity modes, as recently show in Ref.~\cite{cao_strong_2021}. In panel (e) we plot the linewidths of the two polariton modes and of the bare resonances. Contrary to the Lyman-alpha transition, in this case the very small Ps linewidth leads to cavity-dominated polariton linewidths. While only $r=2$ repetitions are already sufficient to observe SC even for the lowest densities considered here, this means that positronium-polaritons could in principle be observed at even lower densities by increasing the repetition number, and thus the quality factor of the DBRs.

\begin{figure*}[tb]
    \centering
    \includegraphics[width=\textwidth]{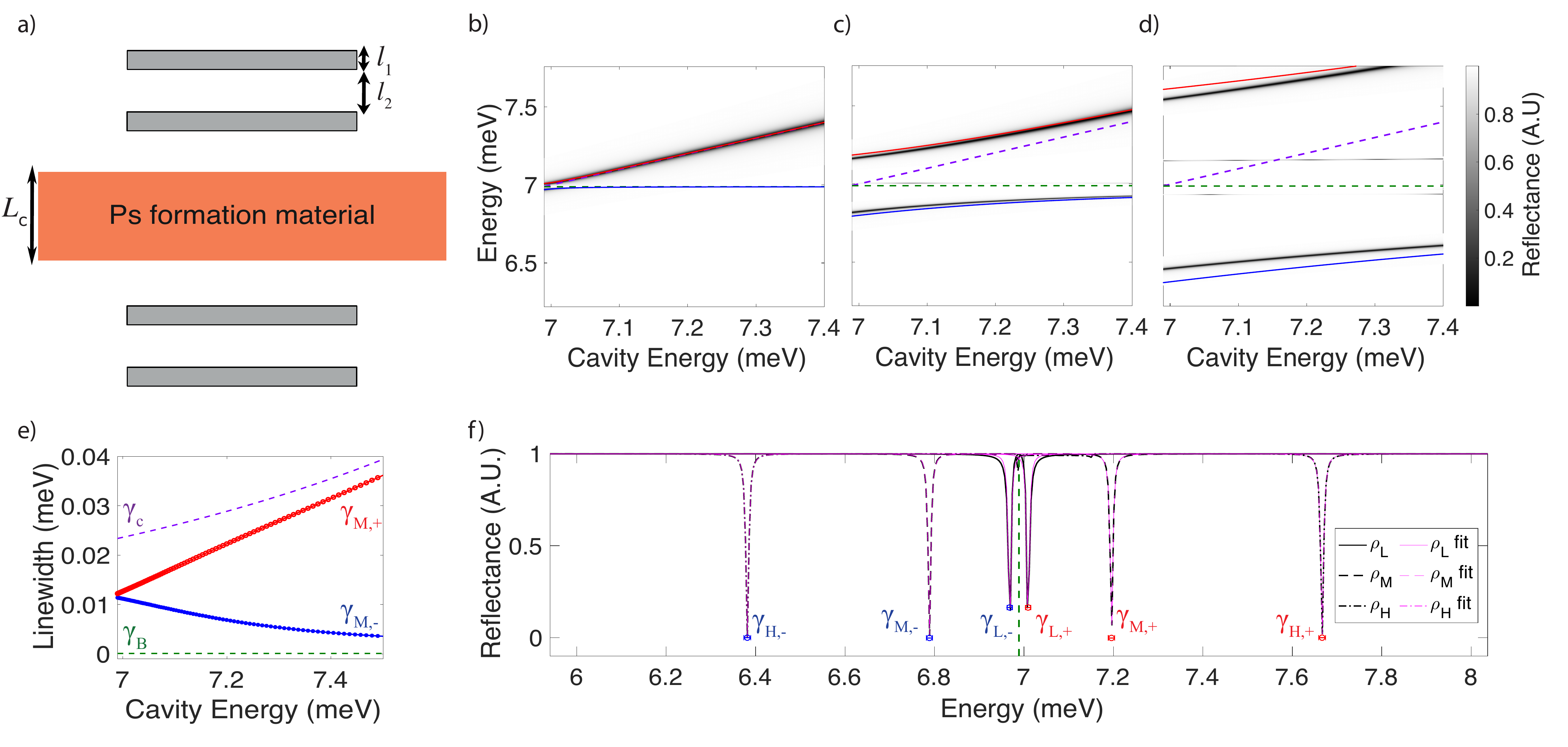}
    \caption{Same as in Fig. \ref{fig2} but for the $12\rightarrow 13$ Rydberg-Rydberg transition, considering a DBR composed of 
    $r=2$ Si layers in air. The used densities are one order of magnitude lower: $\tilde{\rho}_{L}=10^{15}$ cm$^{-3}$ (b), $\tilde{\rho}_{M}=10^{17}$ cm$^{-3}$ (c), and $\tilde{\rho}_{H}=10^{18}$ cm$^{-3}$ (d), in order to take into account the lower efficiency of Rydberg state creation.}
    \label{fig4}
\end{figure*}

From these results it appears that the parameters required to achieve SC using Rydberg states are much less stringent than in the previous case, although this comes at the cost of the extra complexity required to create the Rydberg states themselves. 

\section{Density measurement}
\label{sec:designs}

In Fig. \ref{fig6}(a) we plot the expected resonant VRF $\Omega_A$ as a function of the Ps density and highlight in yellow the parameter region of SC, where the Rabi splitting overcomes the dominant effective loss rate $\gamma_{A}$.
We focus here on the 1S~$\rightarrow$~2P Lyman-alpha transition because, owing to the the smaller dipole moments and shorter lifetimes, this case is more strongly affected by the density restrictions. The errors on the polariton splittings (blue horizontal errorbars), derived from the polariton resonances linewidths, propagate onto the estimated Ps density, leading to the uncertainty values marked by magenta vertical errorbars. It is clear that the measurement becomes more accurate in the high density range. 

We mark with coloured symbols four densities achieved or predicted in various experimental geometries, which we will now briefly describe. Ps produced in mesoporous silica films may have lifetimes that are a significant fraction of the intrinsic vacuum lifetime, depending on pore size~\cite{gidley_positronium_1999}. Positrons implanted into such materials form Ps atoms which then migrate towards the voids of the substrate's typical ``swiss cheese" pore structure, where they may thermalize and become trapped. The maximum density achieved so far using this method is $\rho\approx 10^{16}$~cm$^{-3}$~\cite{cassidy_efficient_2012} [see Fig. \ref{fig6}(b)]. This density can be increased using higher incident positron beam densities obtained, for example, by brightness enhancement techniques~\cite{Mills1980}.  

Several different Ps production arrangements have been suggested to achieved higher Ps densities: Mills has suggested using a collection of small disconnected cavities made by seeding a porous silica target film with monodisperse $20$~nm diameter polyethylene porogen particles [see Fig.~\ref{fig6}~(c)] to produce Ps densities on the order of $\rho=3\times10^{19}$ cm$^{-3}$~\cite{Mills2019}. Another proposed design, shown schematically in Fig.~\ref{fig6}~(d), features a ``straw hat-shaped" cavity, in which Ps atoms would gather after thermal diffusion through the porous silica target. This arrangement is predicted to result in a Ps density of $\rho=2\times10^{18}$ cm$^{-3}$~\cite{Mills2019}. To deal with heat dissipation the authors of Ref.~\cite{asaro_conditions_2022} have proposed a cavity fabricated at the interface between a porous silica substrate and a thin layer of isotopically pure diamond single crystal, where implanted positrons may cool down before forming Ps atoms (Fig. \ref{fig6}(e)) The authors estimate that the final Ps density within the cavity walls will be $\rho\approx4\times 10^{19}$ cm$^{-3}$. Finally, Mills has also proposed a totally different approach, in which Ps atoms may be formed in bubbles of superfluid liquid $^4$He, and has suggested that densities of $\rho\approx 10^{20}$ cm$^{-3}$ may be achievable~\cite{mills_positronium_2019}.

\begin{figure}[tb]
    \centering
    \includegraphics[width=0.5\textwidth]{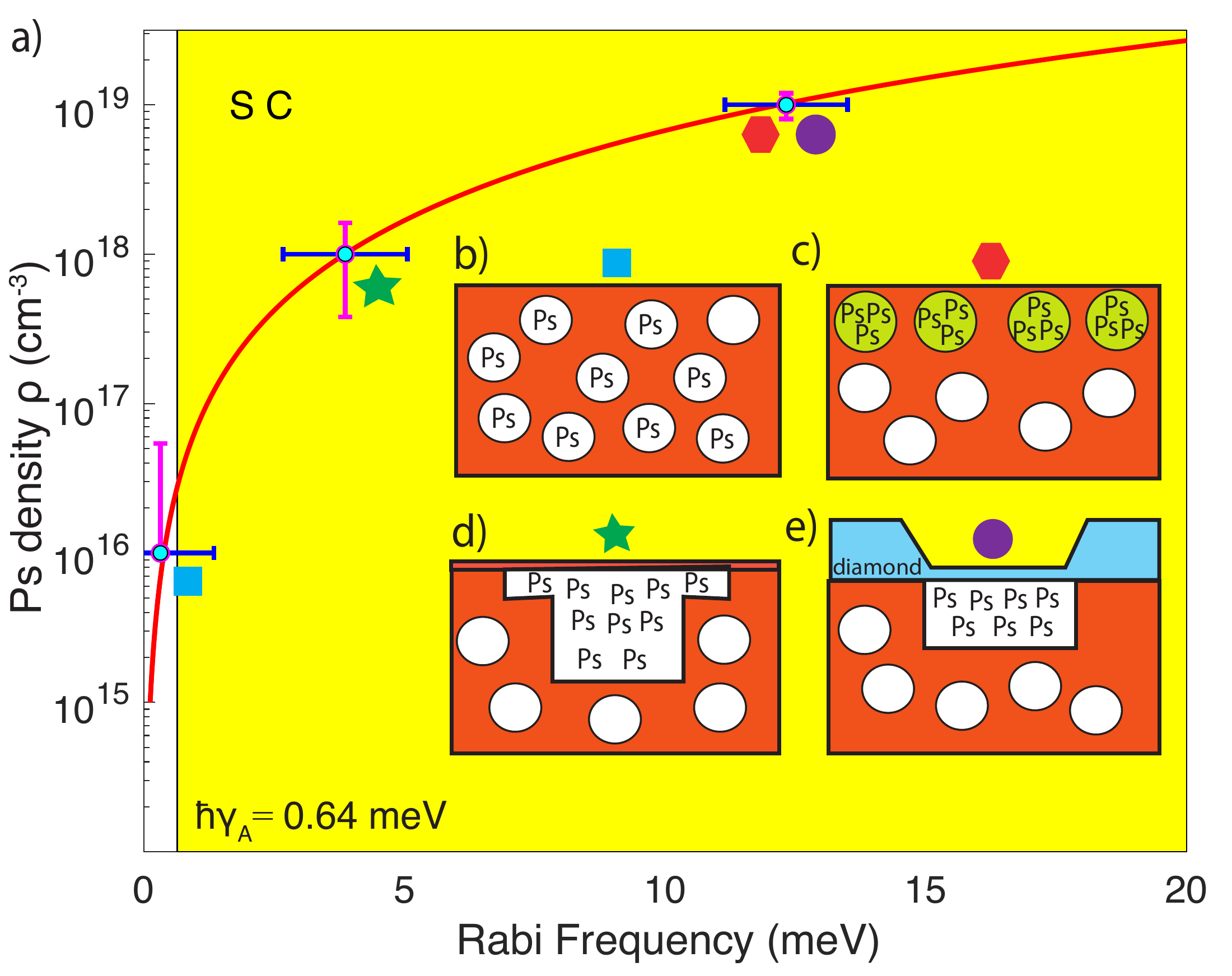}
    \caption{
     (a) Plot of the relation in Eq. $\ref{OR}$ between VRF and Ps density. The area achieving SC is shaded in yellow. Cyan dots represent the values of the VRF derived from the numerical simulations in Sec. \ref{subsec:1S}, with blue horizontal errorbars, describing the uncertainties on the splitting measure, and vertical magenta errorbars representing the propagated error on the Ps density. (b-e) Sketches of the different porous silica target geometries described in the text. The four coloured symbols mark the Ps density achieved or expected for each one.}
    \label{fig6}
\end{figure} 
 
\section{Conclusions}
In this paper we explored the possibility of strongly coupling a high-density Ps atomic gas to a single-mode photonic resonator, investigating dipole-allowed transitions from both the ground state and from highly excited states, thus addressing completely different energy scales. We find that positronium-polaritons may be achievable using existing positron beam technology in the far-infrared or ultraviolet regimes. We proposed that the collective nature of light-matter interaction in CQED can thus be used to provide an ultra-fast spectroscopic measurement of the Ps density evolution inside a cavity in density regimes of relevence to the formation of a Ps BEC. We also demonstrated that the ultrastrong light-matter coupling regime could be achieved in positronium-polaritons, introducing a novel platform to explore its non-perturbative phenomenology with potentially unparalleled accuracy.

\section*{Acknowledgements}

S.D.L. acknowledges the support of a University Research Fellowship of the Royal Society and of the Phillip Leverhulme prize of the Leverhulme Trust. S.D.L. and E.C. acknowledge funding from the RPG-2022-037 grant of the Leverhulme Trust. D.B.C acknowledges support from EPSRC via grant~EP/S036571/1. 

\appendix 

\section{Calculation of polariton modes via bosonic diagonalization} \label{sec:bosonic}

The interaction between an optically active transition of frequency $\omega_t$ in a gas of $N\gg 1$ Ps atoms and a single-mode cavity field oscillating at frequency $\omega_c$, can be described in terms of two coupled harmonic oscillators, even for inhomogeneous Ps distributions \cite{Gubbin2016}. The phenomenological Hamiltonian can be written in the Rotating Wave Approximation (RWA) as 
\begin{eqnarray} \label{H}
H= \hbar\omega_c a^\dagger a +\hbar \omega_t b^\dagger b + \Omega_t \left( a b^\dagger+ a^\dagger b \right),
\end{eqnarray}
with $a$ and $b$ the annihilation operators for the photonic and matter excitations respectively, obeying the bosonic commutation relations $\left[a,a^\dagger\right]=\left[b,b^\dagger\right]=1$, and $\Omega_t$ the VRF from Eq.\ref{OR}.
The Hamiltonian in Eq. \ref{H} can be diagonalised by introducing the polaritonic  operators
\begin{eqnarray}
p_j= x_j a  + y_j b,
\end{eqnarray}
where $j=\pm$ indexes the two polariton branches, and $x_j$ and $y_j$ are the Hopfield coefficients obtained by solving the eigenvalue equation for the polariton frequencies $\omega_{j}$
\begin{eqnarray} \label{eig}
\hbar \omega_{j} p_j= \left[p_j,H\right].
\end{eqnarray}
The polaritonic frequencies can be written
\begin{eqnarray} \label{eig_pol}
\omega_{\pm}= \frac{1}{2}\left[\omega_c+\omega_t \pm \sqrt{4 \Omega_t^2 +\left(\omega_c-\omega_t\right)^2}\right],
\end{eqnarray}
that, in the resonant case $\omega_c=\omega_t$, leads to a polariton splitting 
\begin{eqnarray} \label{spl}
\omega_{+}-\omega_{-}=2 \Omega_t.
\end{eqnarray}

\section{Transfer Matrix Approach} \label{sec:TM}

The optical reflectivity spectra of a multilayer planar Ps-material-cavity system can be predicted via transfer matrix simulation \cite{kavokin_microcavities_2017}.
We can model the system as a series of planar layers, including those composing the cavity mirrors and the Ps host material.
Considering the probe electric field propagating normally to the interface plane, the transfer matrix $T_j$ across the $j$th layer of thickness $l_j$ and frequency-independent refractive index $n_j$ can be written as 
\begin{eqnarray} \label{TM1}
T_j=  \left(
 \begin{array}{cc}
   \cos{(k_j l_j)} &   i\sin{(k_j l_j)}/n_j\\
    i\sin{(k_j l_j)} n_j  &  \cos{(k_j l_j)} 
 \end{array}\right),
\end{eqnarray}
where $k_j=\omega n_j/c$ with  $c$ the speed of light and $j$ the layer index.
We can model the dielectric function of light interacting with the transition at frequency $\omega_t$ in the Ps medium using a Lorentz dielectric function \cite{cao_strong_2021}
\begin{eqnarray}
   \epsilon_{t}(\omega)=\epsilon \left(1-\frac{L_{\textrm{eff}}}{L_{c}}\frac{2\Omega_{t}^2  }{\omega^2-\omega_{t}^2+i\gamma_{t} \omega}\right),
\end{eqnarray}
where $\gamma_{t}$ is the effective Ps radiative decay rate for the considered transition.  

The total transfer matrix can then be expressed as the product of the matrices for each layer
\begin{eqnarray}
T= \Pi_j T_j.
\end{eqnarray}
As an example, if we consider the optically active material embedded between a pair of DBRs, with $r$ AlGaN/AlN alternating layers, the total transfer matrix can be written
\begin{eqnarray}
T= \left(T_a\cdot T_b\cdot \right)^{r}\cdot T_c\cdot T_d \cdot \left(T_a\cdot T_b\cdot \right)^{r},
\end{eqnarray}
with $a,b,c,d$ corresponding respectively to the AlN and AlGaN layers of the DBRs, the spacing medium between the two mirrors and  the Ps host material.
The reflectivity $R(\omega)$ of the whole stack embedded between two semi-infinite media (air) of unitary reflective index in terms of the transfer matrix elements can finally be written as 
\begin{eqnarray} \label{R}
R(\omega)=\left | \frac{T_{1,1}+T_{1,2}-T_{2,1}-T_{2,2}}{T_{1,2}-T_{1,1}+T_{2,1}-T_{2,2}} \right|.
\end{eqnarray}

\bibliography{Ps_zot2}

\end{document}